# Thermal Instability and Current-Voltage Scaling in Superconducting Fault Current Limiters

B. Zeimetz[1,3#], K. Tadinada[2], D. E. Eves[2], T. A. Coombs[2,3], J. E. Evetts[1,3] and A. M. Campbell[2,3]

[1]Department of Materials Science and Metallurgy, Cambridge University, Pembroke Street, Cambridge CB1 3QZ, United Kingdom
[2]Department of Engineering, Cambridge University, Trumpington Road, Cambridge, .U.K.
[3]IRC in Superconductivity, Cambridge University, Madingley Road, Cambridge, U.K.

## Abstract

We have developed a computer model for the simulation of resistive superconducting fault current limiters in three dimensions. The program calculates the electromagnetic and thermal response of a superconductor to a time-dependent overload voltage, with different possible cooling conditions for the surfaces, and locally variable superconducting and thermal properties. We find that the cryogen boil-off parameters critically influence the stability of a limiter. The recovery time after a fault increases strongly with thickness. Above a critical thickness, the temperature is unstable even for a small applied AC voltage. The maximum voltage and maximum current during a short fault are correlated by a simple exponential law.

## 1. Introduction

Superconducting Fault Current Limiters (SFCL) are one of the most attractive applications of High Temperature Superconductors in electrical power engineering [1]. They could replace the slow and resistive mechanical switches which are being used in high/medium voltage transmission and distribution networks, thereby enhancing power quality and network efficiency. Two distinctive design concepts have evolved for SFCL: Resistive current limiters utilise the nonlinear resistivity of superconductors, while inductive SFCL are based on their magnetic shielding properties.

In order to design SFCL for practical applications, knowledge of SFCL parameters such as maximum load, response time and thermal recovery time is essential. On the other hand, the electromagnetic and thermal response of a SFCL to a fault involves very high voltages and currents at very short times, and is therefore a formidable challenge for experiments. Computer simulations are obviously helpful, since they allow research on arbitrary time scales and power levels.

This article presents recent results of SFCL simulations. In line with recent and current SFCL developments in Cambridge [2], it is focussed on resistive SFCL. The simulation software is based on earlier simulation work at Cambridge University [3][4], which recently has been significantly extended.

## 2. Computer Model

Our simulation uses a finite-difference approach to calculate the electrical and thermal evolution of a SFCL. The superconducting material is modelled electrically as a tetragonal network of nonlinear resistors, and thermally as a three-dimensional array of tetragonal cells, cf. figure. 1. Each cell contains 6 resistors branching out from its centre. The global current $I_{FCL}$ is flowing in x direction, and its magnitude is calculated as function of an applied AC voltage $V_{FCL}$. The current-voltage characteristic of a resistor is defined as

---

# e-mail: bernhard.zeimetz@inpg.fr; present address: Laboratoire des Materiaux et du Génie Physique, Institut National Polytechnique de Grenoble, F-38402 St. Martin d'Hères, France

$$V_R = 0 \qquad\qquad \text{for } I < I_{CRi} \qquad\qquad\qquad (1a)$$

$$V_R = V_{0R}*(I/I_{CRi} - 1)^n \quad \text{for } I > I_{CRi} \qquad\qquad\qquad (1b)$$

where $V_{0R}$, $I_{CRi}$ ($i$ = xy-plane or z-axis) and n are parameters defined locally for each cell, and as function of temperature and magnetic field. This allows the study of inhomogeneous, anisotropic superconductors.

The calculation of the current distribution involves a Newton-Raphson iteration to minimise the resistor network loop voltages as function of the loop currents [5]. By explicitly utilising the tetragonal symmetry of the model, the calculation speed can be dramatically increased, and the required memory size minimised, so that even large model sizes can be tackled on a standard personal computer. (For example, the calculation shown in Fig. 7 with 2000 cells and 1000 time steps takes ca. 150 min on a 400 MHz PC).

The thermal response of the FCL to the current is governed by locally defined, temperature dependent specific heat $c_V$ and anisotropic thermal conductivity $\lambda$. For the boundary conditions, i.e. the heat flow through the surfaces, one can choose between thermal insulation, thermal sink and boil-off of a cryogenic liquid. The heat transfer properties of the boil-off model are shown in figure 2. The data for the boil-off power were estimated from experimental data given in Ref. [6]. These experimental data scatter over a wide range, up to an order of magnitude. The boil-off data are sensitive for example to the sample geometry and its surface roughness. Therefore our boil-off model can only be seen as a rough estimate, unless boil-off power for a specific sample can be measured.

The second important feature of the nitrogen boil-off model shown in figure 2 is the existence of a maximum in the temperature dependence of the boil-off power. This leads to thermal instability, as will be discussed below. The maximum originates from the behaviour of the nitrogen gas on the surface, which at low temperatures exists as individual bubbles ('nucleation boiling'), but then evolves into a continuous gas layer with poor heat transfer ('film boiling') [6].

Using $Bi_2Sr_2CaCu_2O_8$ (Bi2212) as a model material for our studies, we extracted the electrical input parameters $V_{0R}(T,B)$, $I_{CR}(T,B)$ and $n(T,B)$ from our own measurements on Bi2212 polycrystalline CRT bars [7][8] carried out in a pressurised liquid nitrogen vessel [9][10]. However, reliable c-axis critical current data were not available. Owing to the imperfect texture of the Bi2212 grains in CRT bars, we estimated the critical current anisotropy to be 10 and used this value for our simulations. Thermal data for Bi2212 were extracted from measurements on single crystals [11-13]. In this paper we consider mainly slab-like conductors modelled on CRT bars with length *L*, width *W*, thickness *D*, and *D* < *W* < *L*, see Fig. 1.

## 3. Results

### 3.1 Thermal instability in normal (non-fault) operation

The plateau and subsequent decrease in the boil-off heat transfer at the surface leads to thermal instabilities or meta-stable states. The most straightforward case is that of a DC current applied to an ohmic conductor with temperature-independent resistivity. The simulated temperature-time dependence of this scenario is shown in Fig. 3, for different voltages. At low voltages, the conductor reaches an equilibrium temperature at which the heat generated by the electrical current is equal to the boil-off power at the surface. If however the temperature exceeds the plateau onset ($T_{B1}$) of the boil-off function ($T_{B1}$ = 89 K in our model, cf. figure 2), no such equilibrium exists, and we observe a thermal 'runaway' of the conductor which in an experiment would lead to its destruction.

We have also carried out simulations with a more realistic temperature-dependent resistivity. If the resistivity increases with temperature, the thermal instability is somewhat mitigated, but the qualitative behaviour is very similar. Similarly, replacing the DC with an AC voltage only results in an additional temperature oscillation around the equilibrium values, but otherwise no change in stability behaviour.

In the case of superconducting materials, the situation is complicated by the strong temperature dependence of the electrical properties, and especially by the existence of the critical temperature $T_c$. Figure 4 shows our DC simulation for Bi2212 material, where the value of $T_c$ is 90 K and therefore very close to the boil-off plateau temperature ($T_{B1} = 89$ K). (The vicinity of the two temperatures makes it even more crucial to obtain reliable experimental data for the boil-off parameters). As in the ohmic case, at low DC voltages the temperature converges towards an equilibrium value. If the voltage is raised so that the temperature just exceeds $T_{B1}$, the temperature rises to $T_C$ and then oscillates around this value. When the DC voltage is increased further, the temperature diverges as in the ohmic case.

The simulations discussed above and shown in figures 3 and 4 were based on a two-dimensional network model, with only a single layer along the z-axis. This allowed us to focus on the boil-off properties and neglect the internal heat flow inside the conductor. However, the thermal conductivity in Bi2212 is lower by a factor of 100 compared to conventional metals, and it is also anisotropic with a lower value along the crystallographic c-axis [12]. Therefore, in fully three-dimensional simulations, we find a markedly higher temperature gradient in simulations with Bi2212 material, and we also find a strong dependence of thermal behaviour on sample thickness. Figures 5 and 6 show simulation data from a fully three-dimensional calculation, in this case using AC applied voltage. The bottom z plane was thermally insulated, and it therefore represents the centre of a symmetric slab of dimensions $L \times W \times D = 20 \times 10 \times 4$ mm. All other surfaces were cooled by nitrogen boil-off. Here the surface temperature converges towards a value below $T_c$, while the core temperature is heated above $T_c$. Apart from the oscillations caused by the AC current variation, shown in the inset, the temperature distribution reaches a stable, partially quenched state. Figure 6 shows the temperature distribution of the central layers in the 'equilibrium' state as a grey scale, and the corresponding local current distribution, indicating local current density by the length of the arrows. The current flows predominantly along the surface layers which are still superconducting. We have verified that this result is independent of the number of z-layers used, i.e. it is not a numerical effect of the discretisation.

Similarly to the one-layer DC case (figure 4), we found that the equilibrium temperature distribution depends on the applied voltage. More interestingly, the equilibrium temperature distribution also depends on the thickness $D$ of the slab: Because the surface-per-volume ratio decreases with increasing thickness, the equilibrium temperature increases with $D$. This is shown in figure 5 which includes also the temperature-time data of a simulation with thickness D = 0.2 mm (short line near bottom), at the same voltage 20 mV and same other parameters. In this case, all 5 layer temperatures practically overlap, and the temperature gradient is very small.

We can conclude that, above a critical thickness of the superconductor, even a small voltage leads to a (partial or complete) quench of a SFCL after a few seconds. While this is not very relevant for quench protection in SFCL itself, which can be backed up by mechanical circuit breakers, a low level of dissipation is inevitable in AC applications due to AC losses. This dissipation might be sufficient to trigger a partial quench during normal, i.e. non-fault operation in SFCL consisting of thick slabs or other bulk superconductors, making them inherently unstable. In the next section we show that also the thermal recovery time after a fault strongly depends on the thickness.

## 3.2 Dynamical Response and Post-Fault Recovery Time

We now consider the behaviour of a SFCL during and after a fault. Our 'model' fault consists of a short high voltage peak, typically 10 ms long and several volts high. An example is shown in figure 7. Before and after the fault, a low 'base voltage' of 20 mV/50 Hz AC is applied. (Due to numerical problems, we cannot simulate a truly superconducting state with zero voltage and finite current) Also shown in figure 7 are the overall current through the SFCL, and the maximum temperatures of the layers along z-axis. The highest temperature is reached in the core, which also is slowest to recover to its equilibrium temperature.

One of the most important characteristics of a SFCL is its recovery time after a fault. We define the recovery time $t_R$ as the time at which the maximum temperature is reduced to 1/e of the difference between peak and equilibrium temperatures, as indicated in the inset to Fig. 8. Using this definition, we have calculated the recovery time as function of slab thickness, and the result is shown in the main plot of figure 8. The calculated result lies somewhat in between linear and quadratic dependencies, which are indicated with dashed and dotted lines, respectively.

A quadratic dependence can be expected if the heat transfer is confined to one direction (in our case the z-axis, if D << W, L), and if the heat is generated only in one plane $z_0 = 0$. This problem can be solved analytically [14], and we get

$$t_R = (D/2)^2 * c_V / \lambda_C \qquad (2)$$

where $\lambda_C$ is the thermal conductivity along the c-axis. This is the $t_R(D)$ dependence plotted as a dotted line in figure 8 (using $c_V(77K)$ and $\lambda_C(77K)$). However, this model with heat source at centre is obviously is too simplistic to explain our simulation data, because in the simulation the ohmic heating is spread across the entire volume. Nevertheless, the thickness is clearly a crucial factor determining the recovery time of a SFCL.

### 3.3 Observation of 'hot spots' at Contacts and Inhomogeneities

Up to this point we have only discussed SFCL simulations of homogeneous superconductors and metals. However, in our model we can define materials parameters locally for each cell, and thereby study effects of inhomogeneity, contacts and geometric constrictions.

A simple example is shown in figure 9. Here a defect was defined near the central part of a Bi2212 slab. In the defect volume, which is marked by the black rectangles, the critical current density was set to be reduced by 50 %, see equation 1. The current flows around the defect as expected. Perhaps more surprising is the temperature distribution, which is shown in grey colour scale: The hottest point is not *at* the defect, but next to it, where the current density is highest.

The metal contacts of a Bi2212 CRT bar consist of silver sheets which are partially embedded in the superconducting bar. After a heat treatment, this results in very low contact resistance [8]. The open part of the silver sheet can be clamped to current leads. This contact geometry was modelled in our simulation as shown in figure 10a. The arrows in figure 10a represent a typical current distribution in normal SFCL (non-quench) mode ($V_{FCL.}$ = 20 mV AC). One can see that the largest part of the injected current is transferred to the superconductor at the end of the bar. This finding coincides with the isothermal finite element calculations carried out by Kursumovic et al. [8].

Figure 10b shows the current and temperature distribution at the onset of a quench ($V_{FCL.}$ = 1 V AC). In this case, a large part of the injected current flows through the entire length of the silver contact. As a result, a 'hot spot' forms in the superconducting material just behind the end of the silver sheet. A second, weaker hot spot is present in the bottleneck of the metal sheet just outside of the Bi2212 bar. This suggests that by careful design of the contacts, hot-spot formation can be controlled and mitigated.

### 3.4 Current - Voltage scaling

The time development of voltage, current and temperature in a simulation using a CRT model geometry is very similar to that in a homogeneous slab as was discussed above and shown in figure 7. We can observe in figure 7 that the peaks in voltage, current and temperature all occur at different times. It is therefore surprising that the peak voltage and peak current are related by a simple exponential law, as is shown in figure 11 for a CRT bar, and described by equation 3:

$$V_{p,FCL} = V_{p0} * \exp(I_{p,FCL}/I_{p0}) \qquad (3)$$

This result was confirmed for a number of different parameter sets, and in particular also for inhomogeneous materials, similar to that shown in Fig 9. If this exponential form is indeed a universal one, which has to be confirmed experimentally, the current-voltage response of a SFCL would be determined by only two parameters $V_{p0}$ and $I_{p0}$, which in turn could be found from only two measured data points. This would significantly simplify design considerations. It should also be noted that equation (3) is valid independently of whether the highest temperature $T_{p,FCL}$ is below or above the superconductor's critical temperature $T_c$. This temperature is usually reached at a later time compared to $V_{p,FCL}$ and $I_{p,FCL}$.

However, equation (3) does not predict the destruction point of the SFCL, i.e. the highest voltage which can be applied before the SFCL suffers permanent damage. In our simulation, we defined this point rather arbitrarily, by defining a maximum possible temperature at 300 K. The peak temperature is plotted against peak current in figure. 12. $T_P(V_p)$ first increases gradually up to $T_c$, where it reaches a plateau, but then increases very rapidly. This functional behaviour can be explained by the variation of boil-off efficiency shown in figure 2 and discussed in section 3.1. We found that the point at which the destruction temperature is reached depends critically on the input parameters. A more quantitative investigation is planned for the future.

## 4. Conclusions

Our simulations of normal (low-voltage) operation and quench in SFCL show that a number of parameters have a critical influence on the stability of SFCL: The boil-off efficiency is strongly temperature dependent, which can lead to instabilities. The thickness of a resistive SFCL strongly influences its recovery time, and thick slabs can be thermally unstable even at low voltages. 'Hot spots' form predominantly near defect points, but also close to embedded metal contacts. The maximum voltage and current reached during a short fault are correlated by a simple exponential law.

The results of our simulations so far are mainly of qualitative nature, due to the uncertainty in some of the input parameters, namely the boil-off efficiency function and c-axis critical currents. An urgent task for future work is therefore to obtain more reliable data. This should then allow us to make quantitative predictions of SFCL behaviour, and to test these against experimental results from Bi2212 CRT bars and other materials.

## Acknowledgments

We thank D. Klaus and P. Barnfather for many valuable discussions, and Dr. J Loram for generously supplying specific heat data of Bi2212. This work was funded by VA TECH T&D and the U.K. Department of Trade and Industry

# Figures

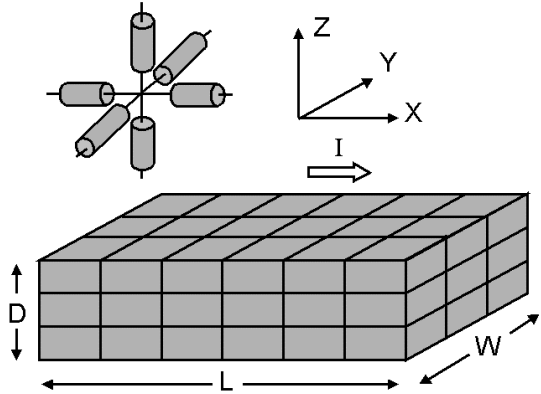

Fig. 1: schematic representation of the model geometry and resistor network; top left: resistors in an individual cell branching out from the centre; bottom: tetragonal cells form a slab-like geometry with applied voltage and current along x direction

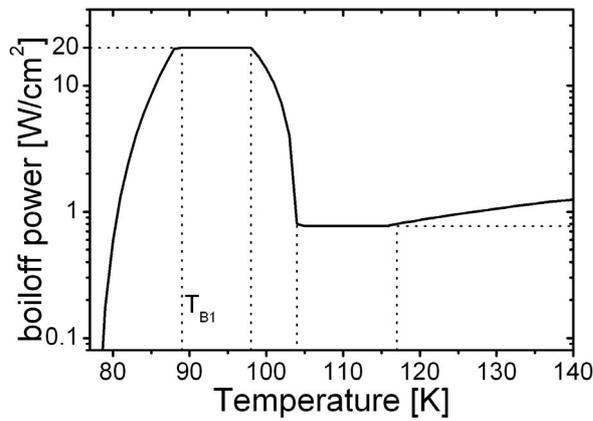

Fig. 2: nitrogen boiloff power per surface area against temperature, estimated from [1], as used in the simulation for cooling at surfaces of SFCL; dotted lines indicate transition temperatures and plateau heights; note the semi-logarithmic scaling

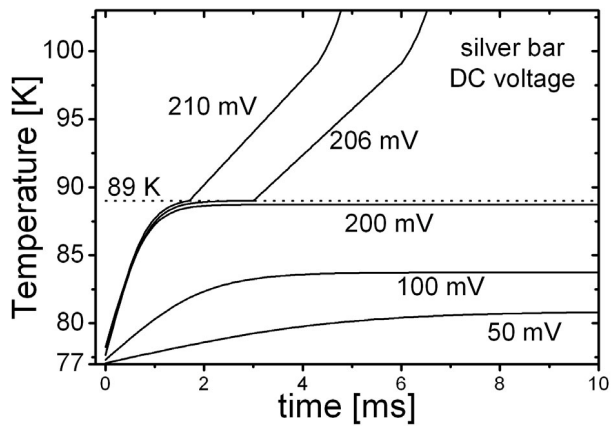

Fig. 3: simulated temperature against time of a silver bar, with various applied DC voltages as indicated, dotted line indicates temperature of 89 K, where the bar temperature becomes unstable (cf. Fig 2); a second upturn can be seen at around 98 K

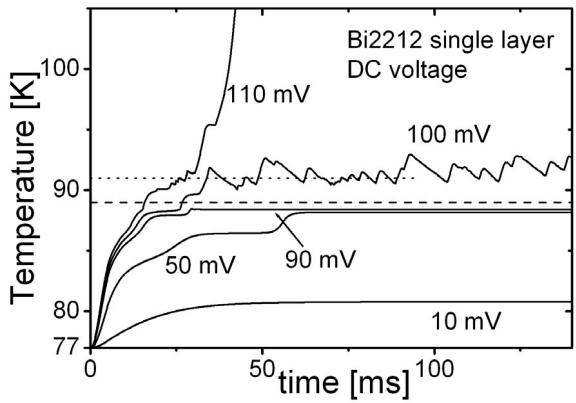

Fig. 4: simulated temperature against time of a Bi2212 bar, with various applied DC voltages as indicated, dotted and dashed lines indicate temperatures of 89 K and 91 K, respectively

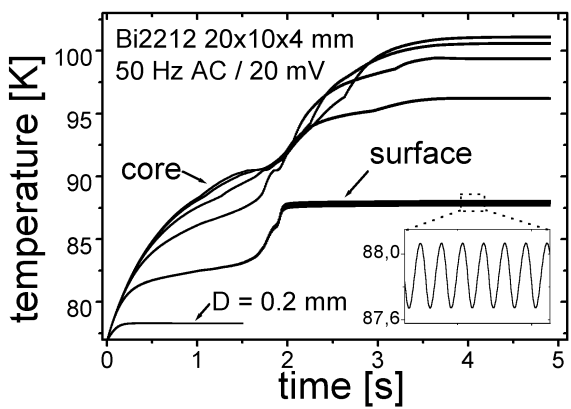

Fig. 5: temperature-time dependence calculated in a simulation of a Bi2212 slab using 5 layers along z; showing temperature maxima in each layer; inset: zoomed view of surface temperature showing oscillation; line marked 'D = 0.2 mm' is a simulation with this thickness, here all 5 temperatures practically overlap

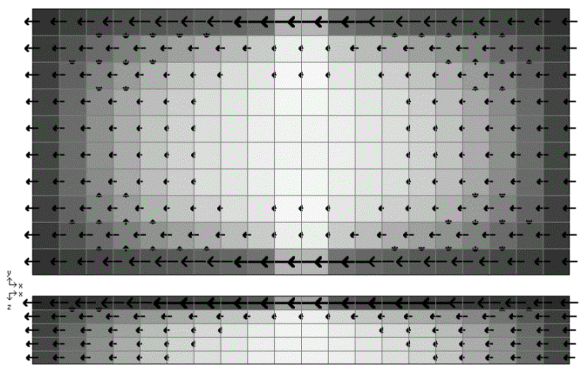

Fig. 6: temperature and current distributions in a simulated Bi2212 slab (LxWxD = 20x10x4 mm) after 5 sec at 20 mV / 50 Hz AC (cf. Fig 5); top: view on bottom x-y layer; bottom: side view on central x-z layer; grey scale represents temperature between 77 K (black) and 92 K (white); arrows represent currents with arrow length and thickness indicating current density

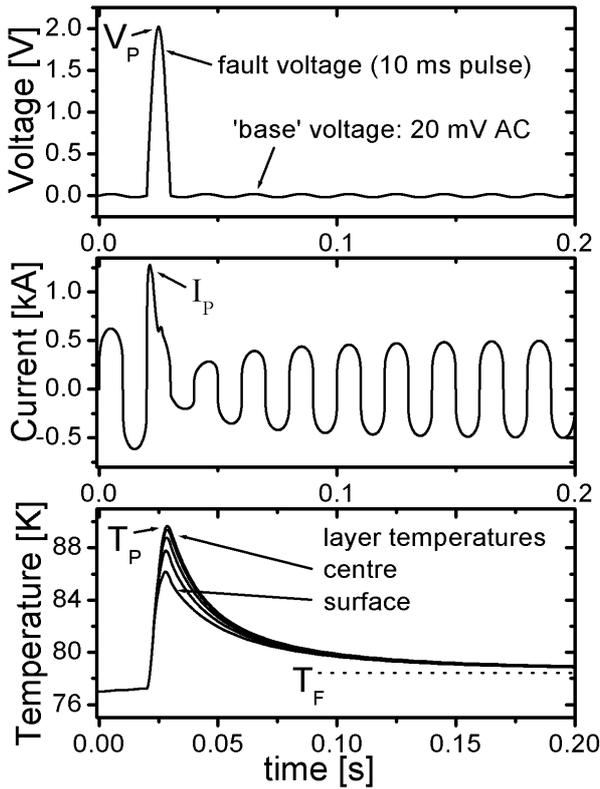

Fig. 7: SFCL model calculation using a homogeneous bar with Bi2212 material data. The applied voltage $V_{FCL}$ (t) is the simulation input, in this case using a 50 Hz base voltage of 20 mV, and a sinusoidal pulse (50 Hz, 10 ms) to simulate a fault. The SFCL current $I_{FCL}$(t), and max. layer temperatures are calculated as function of time. Also indicated are peak voltage ($V_P$), current ($I_P$) and temperature($T_P$)

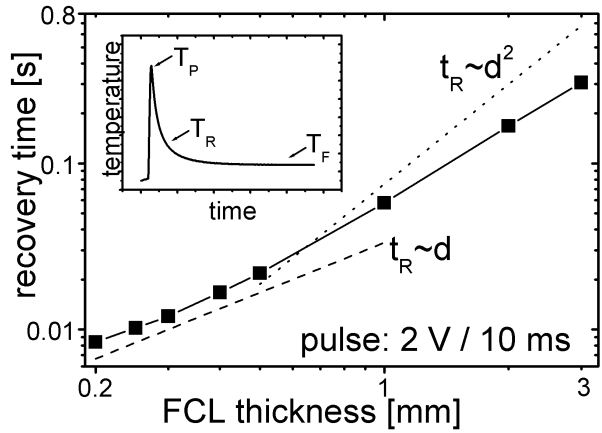

Fig. 8: recovery time against slab thickness, in logarithmic scaling; squares: simulation data for a Bi2212 slab after a 2 V / 10 ms pulse; dashed line: linear dependence; dotted line: quadratic dependence (cf. equation (2) in text); inset: schematic temperature-time dependence with definition of temperatures $T_P$, $T_F$ and $T_R$ (cf text)

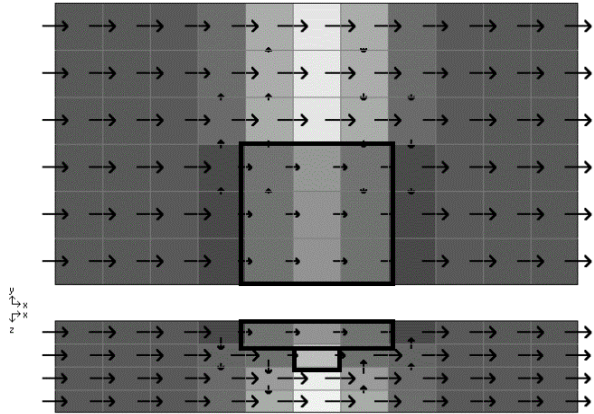

Fig. 9: current and temperature distribution in a Bi2212 slab with a defect volume where $J_c$ is reduced by 50%; top: view from top, bottom: side view; current density indicated by arrow length and thickness, temperature indicated with grey scale from 77 K (black) to 90 K (white); defect volume indicated with thick black line

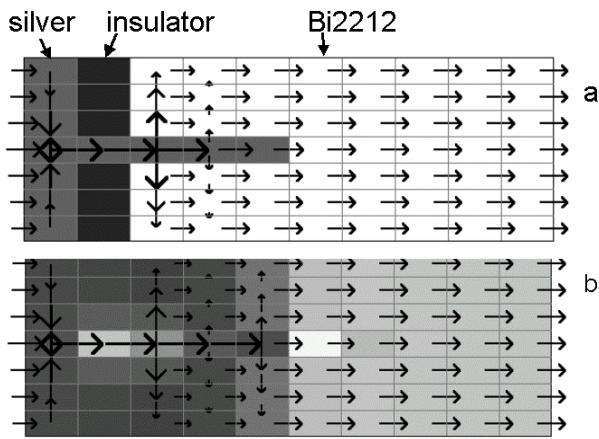

Fig. 10: a) model geometry of the contact area of a Bi2212 CRT bar with embedded silver contacts; in side view, L = 40 mm, D = 2.1 mm (height expanded for clarity); white cells :2212, light grey cells: silver; dark grey: insulating gap between bar and metal clamp (cf. text); arrows: typical current distribution in non-fault operation (V = 20 mV AC) b) current and temperature distribution with applied voltage of 1 V AC, after 25 ms at onset of quench; temperature indicated as grey scale with black = 77 K, white = $T_{max}$ = 81 K; current density indicated by length and thickness of arrows

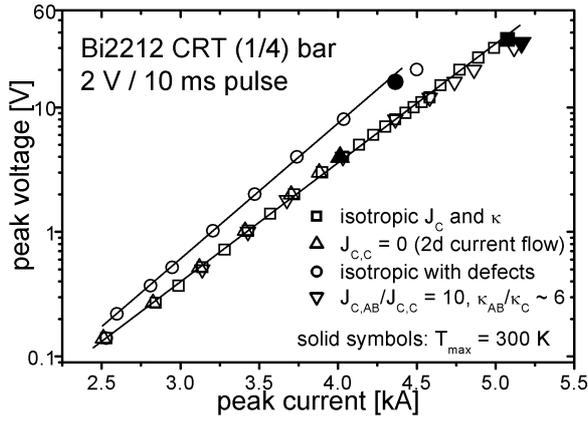

Fig. 11: peak voltage against peak current reached during a 2 V / 10 ms fault in a CRT bar; for different materials parameters as indicated; in semi-logarithmic scaling; full symbols indicate destruction point defined by $T_p$ = 300 K (see Fig 12)

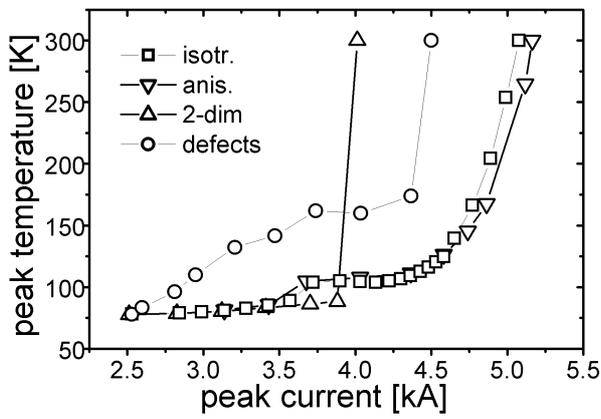

Fig. 12: peak temperature against peak current reached during a 2 V / 10 ms fault in a quarter CRT bar; for different materials parameters as indicated and detailed in Fig. 11